\documentstyle[aps,amssymb,prl,twocolumn]{revtex}

\begin{document}

\newcommand{\bra}[1]{\mbox{$\langle \, {#1}\, |$}}
\newcommand{\ket}[1]{\mbox{$| \, {#1}\, \rangle$}}
\newcommand{\exval}[1]{\mbox{$\langle \, {#1}\, \rangle$}}
\newcommand{\be}{\begin{equation}}
\newcommand{\bel}[1]{\begin{equation}\label{#1}}
\newcommand{\ee}{\end{equation}}
\newcommand{\bea}{\begin{eqnarray}}
\newcommand{\ba}{\begin{array}}
\newcommand{\eea}{\end{eqnarray}}
\newcommand{\ea}{\end{array}}

\draft
\tighten
\onecolumn
\twocolumn[\hsize\textwidth\columnwidth\hsize\csname @twocolumnfalse\endcsname

\title{Annihilating random walks in one-dimensional disordered media}
\author{G. M. Sch\"utz and K. Mussawisade}
\address{Institut f\"ur Festk\"orperforschung, Forschungszentrum J\"ulich,
52425 J\"ulich, Germany}
\maketitle
\begin{abstract}
We study diffusion-limited pair annihilation $A+A\to 0$ on
one-dimensional lattices with inhomogeneous nearest neighbour
hopping in the limit of infinite reaction rate. We obtain a simple exact
expression for the particle concentration $\rho_k(t)$ of the 
many-particle system in terms of the conditional probabilities $P(m;t|l;0)$
for a single random walker in a {\em dual} medium. For
some disordered systems with an initially randomly filled lattice
this leads asymptotically to $\overline{\rho(t)}=\overline{P(0;2t|0;0)}$
for the disorder-averaged particle density.
We also obtain interesting exact relations for single-particle conditional
probabilities in random media related by duality, such as
random-barrier and random-trap systems. For some specific random barrier 
systems the Smoluchovsky approach to diffusion-limited annihilation turns 
out to fail. 
\end{abstract}
\pacs{PACS}

\vskip2pc]

Stochastic reaction-diffusion processes play an important role in the
description of interacting many-particle systems in both physics and
chemistry. Usually real systems are much too complex to be
amenable to analytical or even numerical investigation. However, particularly
in the context of critical phenomena, simple toy models may suffice to
determine universal properties correctly and to predict and explain observed
power laws or universal amplitude ratios. 
Hence it is of importance both to
examine the behaviour of such models and to
understand possible relationships between microscopically different processes
and their characterization in terms of universality classes.
In this letter we investigate a model of annihilating random walkers
describing diffusion-limited pair annihilation (DLPA)
of identical particles in inhomogeneous
media. This process describes both chemical reactions where the particles
change their state into an inert reaction product which takes no part in the
subsequent dynamics of the system, or physical reactions where the particles
actually annihilate under the emission of radiation. For a recent review of
experimental and theoretical applications of the ordered model, see Refs.
\cite{Priv,Card96}.
The model is related by a similarity transformation
to the diffusion-limited coagulation process \cite{Kreb95,Schu97} which
describes e.g.
laser-induced exciton dynamics on polymers.
The model also maps to Glauber dynamics for the
one-dimensional
Ising model \cite{Glau63,Racz85,Sant97} and is therefore of interest
in the study of the kinetics of disordered equilibrium systems.
It plays an important role not only for the study of spin-relaxation phenomena
but also for the solution of more intricate problems such as the derivation
of persistence exponents \cite{Derr95} and non-equilibrium steady states
\cite{Chen89,Hinr97}.

The first comprehensive treatment of
homogeneous diffusion-limited annihilation dates back to
Smoluchovsky's classical work in 1917 \cite{Smol17},
but despite renewed strong interest in the 80's and 90's the
experimentally more realistic case of spatially inhomogeneous particle hopping 
rates, e.g. in the presence of quenched disorder, has so far received very
little attention \cite{Oerd96,Card97,Schu97}.
In homogeneous, translationally
invariant environments the particle density decays with a power law which
depends on the dimensionality of the system. Both theoretically and
experimentally one finds $\rho(t) \sim 1/\sqrt{Dt}$ in one
dimension, in agreement with exact results \cite{BG,Torney}.
Interestingly, this result is at variance with the
(dimensionality-independent) mean-field behaviour $\rho(t) \sim 1/(Dt)$
which is correct only in three (and higher) dimensions.
The amplitude is universal in the sense that it depends neither on
the initial density (for random initial conditions) nor on the reaction
rate \cite{Lee94}.
However, if particles are moving in an non-translationally invariant
energy landscape it is not obvious how this will change the decay of the local
or overall particle concentration. 
The physical motivation behind the study of the one-dimensional
case is not only its experimental relevance for polymer physics, but
also its theoretical importance in the understanding of the role of
fluctuations in low-dimensional systems. In both one- and two-dimensional
systems diffusive mixing is inefficient and leads to the building up
of large-scale correlations. Thus the classical mean-field rate equations for
the study of these systems tend to fail and require a more sophisticated
treatment. 

Before going into details, we remind the reader of the essentials of
Smoluchovsky's reasoning. The idea behind this approach \cite{Smol17} is to 
replace in the mean-field
rate equation for the density $\dot{\rho}(t) = - \lambda \rho^2(t)$
the reaction constant $\lambda$ by an effective time-dependent
reaction rate which is proportional to the diffusive current $j(t)$
into an absorbing particle in a background of constant density. In one
dimension where $j(t) \propto \sqrt{D/t}$ this leads (up to the universal
amplitude which cannot be determined from the Smoluchovski argument) to 
$\rho(t) \propto 1/\sqrt{Dt}$. One would like to know whether the
fluctuation-improved mean-field theory of Smoluchovsky which predicts
the correct behaviour of the one-dimensional ordered system
remains valid in the presence of disorder. Naively, one might expect that the
diffusion constant of the pure system would have to be replaced by some
effective diffusion constant of the disordered system, i.e.
$\overline{\rho(t)} \propto 1/\sqrt{D_{eff}t}$ for the disorder averaged 
density.
But clearly this cannot
always work as can be seen in a simple and natural example:
Suppose one investigates DLPA on an ensemble of ordered chains of
varying length $L \leq \Lambda$, modeling e.g. a polymer mixture of
polymer fragments of varying finite length. This is equivalent to taking an 
infinite
chain and place randomly, but with maximal distance $\Lambda$, broken bonds
across which particles cannot move.
If initially particles are placed randomly
with probability 1/2 on each lattice site, then the steady state density in
each chain of length $L$ is equal to $\rho^\ast = 1/(2L)$. Assuming that
each chain length occurs with equal probability, then the averaged density
$\overline{\rho^\ast} = \overline{1/(2L)} \sim \ln{\Lambda}/(2\Lambda)$ for 
large
maximal length $\Lambda$. On the
other hand, since the particle is confined to a box of length $L$ the
system is subdiffusive with $D_{eff}  \propto \Lambda^2/t$. Hence 
the Smoluchovsky formula gives the wrong result $\overline{\rho^\ast} \propto
1/\Lambda$. Even worse, with this Smoluchovsky approach one cannot even 
estimate the approach of the density to its stationary value.
One could, of course, try to be smart and apply Smoluchovsky's approach
to a finite system and then average over system size. This gives indeed
$\overline{\rho^\ast} \propto \ln{\Lambda}/(\Lambda)$. However, if applied
to a different system where one distributes infinitely deep traps (sites out of
which particles cannot jump) at a maximal distance $\Lambda$, then this
refined Smoluchovsky argument would give $\overline{\rho^\ast}  \propto 
\ln{\Lambda}/(\Lambda)$ also for this model. However, the exact result given 
below shows that in fact for this type of disorder $\overline{\rho^\ast} = 
1/\Lambda$ which happens to be consistent with the first, naive Smoluchovsky
result. We conclude that there is no
simple argument which tells one how to proceed without already
knowing the answer. 
Thus, exact results are required for a study of DLPA in inhomogeneous
media.

To this end we investigate by a new exact mapping a  DLPA lattice model
with space-dependent hopping
rates. The particles have no attractive or repulsive interaction
between themselves, they hop with fixed rates $r_k$ ($\ell_k$) from lattice
site
$k$ to site $k+1$ ($k-1$). When two particles meet on site $k$ they both
annihilate
instantaneously. This limit of infinite reaction rate corresponds to the
renormalization group fixed point of the ordered system
\cite{Card96,Lee94,Peli85} and we believe that also the disordered system
with finite reaction rate will be in the same universality class as
the infinite rate limit. 

In the case of constant hopping rates the predictions of this
model are in excellent agreement with experimental data on exciton dynamics
on very long ordered polymer chains. Thus we expect that the disordered model
gives an equally well description of the behaviour of realistic, disordered
systems.

We define the process in terms of a master equation for the probability
$P(\eta;t)$ of finding, at time $t$, a configuration $\eta$
of particles on a lattice of $L$ sites.
Using standard techniques \cite{Kada68,Doi76,Grass} we express the time evolution
given by the master equation in terms of a quantum Hamiltonian $H$.
Since particles annihilate instantaneously when they meet, there can never be
more than one particle on any given site. The idea is now to extend
the bosonic Fock space formalism of \cite{Kada68,Doi76,Grass} to a
Quantum spin chains representation \cite{ST,Alcaraz,Schu95}. One represents each of
the $2^L$ possible particle configurations $\eta$ by a
vector $\ket{\eta}$ which together with the transposed vectors
$\bra{\eta}$ form an orthonormal basis
of a vector space $X=({\mathbb{C}}^2)^{\otimes L}$.
A state $\eta$ with $N$ particles placed on sites
$k_1,\dots,k_N$ is represented
by $\ket{k_1,\dots,k_N}$, the completely
empty lattice by the vector $\ket{0}$.
The probability distribution corresponds to a state
vector $| \, P(t)\, \rangle = \sum_{\eta \in X} P(\eta;t) \ket{\eta}$
and one writes
the master equation in the form
\bel{2-3}
\frac{d}{dt} P(\eta;t) = - \langle \, \eta \, | 
H | \, P(t) \, \rangle
\ee
where the off-diagonal matrix elements of $H$ are the (negative) transition
rates between states and the diagonal entries are the inverse of the
exponentially distributed life times of the states.
A distribution at time $t$ is given in terms of an initial
state at time $t=0$ by $\ket{P(t)} = \mbox{e}^{-Ht } \ket{P(0)}$.
The expectation value $\rho_k(t)=\bra{s} n_k \ket{P(t)}$ for the density at
site
$k$ is given by the projection operator $n_k$ which has value 1 if there
is a particle at site $k$ and 0 otherwise. The vector
$\bra{s} = \sum_{\eta \in X} \bra{\eta}$ performs the average over all
possible final states of the stochastic time evolution.
Choosing the basis of $X$ such that a particle (vacancy) on site $k$
corresponds
to spin up (down) the Hamiltonian
\bel{2-15}
H = - \sum_{k} (r_k h_k^+ + \ell_k h_k^-)
\ee
for the process can be written in terms
of Pauli matrices
$h_k^\pm =(s^+_k s^-_{k\pm 1} + s^+_k s^+_{k\pm 1} - n_k)$
where $n_k = ( 1 - \sigma^z_k )/2$ and
$s^{\pm}_k = (\sigma^x_k \pm i \sigma^y_k)/2$ create ($s^-_k$) and
annihilate ($s^+_k$) particles (see \cite{Schu97} for details).
Since the time evolution conserves particle number modulo 2, it is
convenient to work only on the even and odd subspaces defined by the
projector $P^\pm=(1\pm Q)/2$ where $Q=(-1)^N=\prod_k\sigma_k^z$.
For averaging over final states we then use $\bra{s^\pm}
=\bra{s}P^{\pm}$. The projection on the even sector of the
uncorrelated initial state with a density $1/2$ used below
is given by the vector $\ket{1/2^+}=(1/2)^{L-1}\ket{s^+}$.

In one dimension with {\em homogeneous} nearest neighbour hopping
DLPA is related to zero-temperature Glauber dynamics by a domain-wall duality
transformation \cite{Racz85} which is an invertible similarity transformation 
\cite{Sant97}. On the other hand, zero-temperature Glauber dynamics can be 
brought by another similarity transformation into a form which is the 
{\em transpose} of the Hamiltonian for (homogeneous) DLPA \cite{Henk95}. 
We use these results to construct a new
 matrix ${\cal D}$ such that the process defined by
$\hat{H} =  {\cal D}^{-1}H^T {\cal D}$
also describes a DLPA process with nearest neighbour hopping 
albeit with different hopping rates in the presence of disorder.
We find for the even particle sector
\bel{4-2}
{\cal D}_+ = \gamma_1 \gamma_2 \dots \gamma_{2L-1}
\ee
where $\gamma_{2k-1} = \left[ (1+i) \sigma_k^z
- (1-i) \right]/2$, $\gamma_{2k} = \left[ (1+i) \sigma_k^x\sigma_{k+1}^x
- (1-i) \right]/2$ 
and ${\cal D}_-=-{\cal D}_+\sigma_L^x$
for the odd particle sector. To see what happens under the mapping we
note that ${\cal D}_\pm$ is
unitary and transforms Pauli matrices as follows:
\bea
\label{4-3a}
{\cal D}_\pm^{-1} \sigma_k^x\sigma_{k+1}^x {\cal D}_\pm & = & \left\{
\ba{ll} \sigma_k^z  & k \neq L \\
        Q\sigma_L^z & k = L
\ea \right. \\
\label{4-3b}
{\cal D}_\pm^{-1} \sigma_{k+1}^z {\cal D}_\pm & = & \left\{
\ba{ll} \sigma_k^x\sigma_{k+1}^x  & k \neq L \\
       \pm Q\sigma_L^x\sigma_1^1 & k = L
\ea \right. .
\eea

In the even sector one now finds
\bel{4-4a}
\hat{H} = - \sum_k (r_k h_k^- + \ell_k h_{k-1}^+).
\ee
This process is of the same form as the original
process (\ref{2-15}), but with dual hopping rates
$\hat{\ell}_k = r_k \; , \;\; \hat{r}_k = \ell_{k+1}$.
We shall refer to the environment
defined by the dual rates as to the dual environment \cite{oddsec}. 

In order to make use of the mapping one needs to know how a given initial
distribution and the observables change under the transformation. 
For the transformation laws for states one needs (\ref{4-3a}), (\ref{4-3b})
and
${\cal D}^{-1}_+ \ket{s^+} = -i (i-1)^{L-1} \ket{0}$,
${\cal D}_+ \ket{s^+} = i (-i-1)^{L-1} \ket{0}$ for the even sector
and analogous relations for the odd sector.
For the density at site $k$ for an arbitrary initial state one then
finds
\bel{4-5}
\bra{s^+} n_k e^{-Ht} \ket{P_0} =
\frac{1}{2} \bra{P_0} {\cal D} e^{-\hat{H}t}
(1-\sigma_{k-1}^x\sigma_{k}^x)
{\cal D}^{-1} \ket{s^+}.
\ee
The transformed initial state is a
superposition of the steady state (the empty lattice) and the two-particle
state with particles at sites $k-1,k$.
Bearing in mind that
$\hat{H}$ does not have any particle creation terms one now realizes that the
time-dependence of the density for an arbitrary many-particle initial
distribution
is completely given by the dynamics of just {\em two} annihilating random
walkers in the dual disordered environment, $\rho_k(t) = \alpha-
\sum_{m,l} a_{ml} \bra{m,l} e^{-\hat{H}t} \ket{k-1,k}$ where the
coefficients $\alpha,a_{lm}$ are determined by the initial state
and straightforward to work out \cite{Spou88}.

In order to avoid immaterial technical complications with boundary terms,
we consider from now on only infinite systems. By choosing some of the hopping
rates equal to zero one can always recover results for finite systems.
To calculate the two-particle transition probability
$\hat{P}(m,n;t|k,l;0)=\bra{m,n}e^{-\hat{H}t} \ket{k,l}$ we note that
the transition  probability for a single random walker
$\hat{P}(m,k;t)\equiv\hat{P}(m;t|k,0)$ is the sum over all paths leading from
$k$ to $m$, each
weighted with its proper statistical weight given by the hopping rates and the
particular form of the trajectory. Hence, for two
non-interacting particles moving from $k$ to $m$ and from $l$ to $n$
respectively,
$\hat{P}(m,n;t|k,l;0)=\hat{P}(m,k;t)\hat{P}(n,l;t)$. This sum includes the
contribution of paths which cross each other. In an annihilating random walk
of otherwise non-interacting particles the contribution of all crossing paths
have to be subtracted. Since we are on a one-dimensional lattice
this contribution is just the one given by
all paths which start at site $k$ and end at site $n$ (instead of $m$) and
which start at site $l$ and end at site $m$ (instead of $n$). Therefore
\bel{5-3}
\hat{P}(m,n;t|k,l;0)=\hat{P}(m,k;t)\hat{P}(n,l;t)-
\hat{P}(n,k;t)\hat{P}(m,l;t)
\ee
This further reduces the calculation of the density to the solution of a 
{\em single-particle} random walk
problem in the dual random environment.
For an uncorrelated random initial
state with density 1/2 in the sector of even particle number one gets
\bel{5-1}
\rho_k(t) = \bra{2} e^{-\hat{H}t} \ket{k-1,k}/2
\ee
where $\bra{2}=\sum_{n>m}\bra{m,n}$ is the sum over all states with two
particles. Thus the density at site $k$ is equal to one half the
survival probability of two annihilating
random walkers starting at sites $k-1,k$ and
moving in the dual environment. As a first specific result we calculate
the final density of a system with infinitely deep traps placed randomly
as discussed above.  Clearly, if $k$ is not a trap site, 
then $\rho_k(\infty)=0$. Hence the disorder-averaged density 
$\overline{\rho(\infty)} = 
q \tilde{\rho}(\infty)$ where $q=2/\Lambda$ is the density of traps and
$\tilde{\rho}(t)$ is the density at a trap site. From (\ref{5-1}) one finds
$\tilde{\rho}(t)=1/2$ and therefore $\overline{\rho(\infty)}=1/\Lambda$.

Consider now the relation between the single-particle conditional
probabilities for dual environments. One may write $P(m;t|k;0) =
\bra{s}^{odd} \prod_{i=1}^{m-1}\sigma^z_i n_m \mbox{e}^{-Ht} \ket{k}$.
By taking the transpose and transforming under ${\cal D}_-$ one gets
in the infinite volume limit the interesting exact relation
\bea
P(m,k;t)-P(m,k-1;t) & = & \nonumber \\
\label{10}
\;\;\;\;\;\;\;\hat{P}(m-1,k-1;t)-\hat{P}(m-1,k;t). & &
\eea
Note that eqs. (\ref{4-5}) - (\ref{10}) hold for any fixed hopping
environment, disordered or inhomogeneous, but regular. 
In this letter we focus on disordered systems with translationally and
reflection invariant hopping rate distributions. 
In this case (\ref{10}) gives 
\bel{11}
\overline{P(r;t)}=\overline{\hat{P}(r;t)} + c(t).
\ee
with an undetermined function $c(t)$ which is irrelevant for what follows.
Relations (\ref{10}), (\ref{11}) are remarkable in that they relate the 
conditional probabilities for dual systems.

In order to analyze (\ref{5-1}) further we take a mean field approach
to the disorder average, i.e. we replace in (\ref{5-3}) the disorder
average of the conditional probability for two distinguishable
non-interacting random walkers
$\overline{\hat{P}(m,k;t)\hat{P}(n,l;t)}$ by the factorized average
$\overline{\hat{P}(m,k;t)}\;\overline{\hat{P}(n,l;t)}$.

We have convinced numerically that this factorization holds well
for the random barrier model with uncorrelated bond hopping probabilities 
$b_k$ drawn uniformly from the interval $0.05 < b_k \leq 1/2$. This
was done by exact numerical solution of the discrete-time master
equation for a random walker on a lattice of $L=200$ sites for a given
random realization of the disorder and then taking the average over
$100000$ disorder realizations (to keep disorder-related fluctuations
small). To show this factorization  in Fig. 1 the function

\begin{equation}
\label{num1}
R^{mk}_{nl}(t) = \frac{\overline{\hat{P}(m,k;t)\hat{P}(n,l;t)}}
{\overline{\hat{P}(m,k;t)}\;\overline{\hat{P}(n,l;t)}}-1.
\end{equation}

is shown for $m=n=1$ and $k=l=0$ in a double logarithmic
plot. The return probability gives the largest contribution to  the
function $R^{mk}_{nl}(t)$, hence we do not need to check the
factorization at other positions. The function $R^{10}_{10}(t)$ fitted
well by a power law

\begin{equation}
\label{num1}
R^{10}_{10}(t) \sim t^{-\alpha} 
\end{equation}

where $\alpha = -0.49$. For this the fit routine of the software
package Mathematica was used. We expect that the exact exponent to be
$\alpha = 1/2$. The deviation of the last point in Fig. 1 from the line
originates from finite-size effects. The square root of the time at
this point is about the length of the system and therefor the constant
stationary probability of the random barrier system is reached
exponentially fast. We have also investigated the case where zero is
the lower limit of the distribution of the hopping probabilities in
the same system as above. In this case the behaviour of the mean
square  displacement of a single particle is sub-diffusive $<x^2> \sim
t/\log(t) $ \cite{Alex81,Haus87}. This is due to the existence of very
small hopping probabilities in even very small regions in the chain. The
factorization can only hold for very long times. 
Nevertheless the function $R^{mk}_{nl}(t)$ is very small even for short times
and for longer times there is a tendency of the function $R^{mk}_{nl}(t)$ to
decrease as it can be seen in Fig 2 where $R^{mk}_{nl}(t)$ is spread vs. $log(t)$.
Independent arguments for the validity of this assumption for more general
types of disorder are given in the conclusions.

Then with  (\ref{5-1}) and (\ref{11}) the
disorder average of the density 
for the initially randomly filled 
lattice is given by 

\bel{12}
\overline{\rho(t)} = \left(\overline{P(0;2t)}+\overline{P(1;2t)}\right)/2
\ee
For large times (\ref{12}) becomes the return probability quoted in
the abstract. Having in mind processes like exciton
dynamics it is reasonable to consider (A) random bond disorder 
$r_k=\ell_{k+1}\equiv b_k$
where hopping across a bond $k,k+1$ is symmetric, but bond-dependent
and (B) random site disorder $r_k=\ell_k\equiv s_k$. The energy of a particle 
in the random barrier model (A) is the
same at each site, but between sites there are energy barriers of random height
$E_k$. Thermal fluctuations cause the particle to jump over these barriers 
with a random rate $b_k \propto \exp{(-\beta E_k)}$.
Case B corresponds to the random trap model. Here the particle sits in a
site-dependent potential of depth $-E_k$. Random bond and random site
disorder are dual in the sense of Eq. (\ref{4-4a}). Since for the random-bond
model $P(m,k;t)=P(k,m;t)$  we conclude that the disorder-averaged conditional 
probabilities of the random-barrier model and the
random-trap model are equal in one dimension for any translationally invariant
disorder distribution up to a function $c(t)$.
For an uncorrelated ergodic disorder distribution, diffusion in random barrier
systems converges to Brownian motion, i.e. the averaged conditional probability
becomes asymptotically
equal to a Gauss distribution
with an effective diffusion constant \cite{Alex81,Ansh82,Zwan82,Haus87}
\bel{13}
D_{eff}^{-1} = \overline{b_k^{-1}}
\ee
and therefore asymptotically
\bel{14}
\overline{\rho(t)} =\frac{1}{\sqrt{4\pi D_{eff} t}}
\ee
for random trap and random barrier
systems. For exponentially distributed barrier or trap energies
$\nu(E) \propto \exp{(- E/\sigma)}$ 
the random walk becomes subdiffusive below a critical temperature given
by $\sigma \beta_c = 1$ \cite{Havl,Wich}. This leads to a time-dependent
effective
diffusion
coefficient $D_{eff} \propto t^{(1-\sigma\beta)/(1+\sigma\beta)}$ and hence
to a slower non-universal power law decay of the density $\rho(t) \propto
t^{-1/(1+\sigma\beta)}$.

To conclude, we obtained the following new results. 

(i) We found the duality relations (\ref{10}), (\ref{11}) for random
walkers in dual systems. They play a role in the derivation of
(\ref{12}) - (\ref{14}), but are also of interest in their own right.

(ii) The Smoluchovsky approach to DLPA is consistent with random trap
and random bond systems with vanishing broken bond probability
(\ref{13}), (\ref{14}), but fails for  some important disordered
broken bond systems in one dimension. This raises the question under
which general conditions on the disorder the Smoluchovsky  approach is
correct (see below).  

(iii) The expectation value of the density in diffusion-limited
annihilation in one dimension with nearest neighbour hopping and
infinite annihilation rate is completely determined by the dynamics of
a single random walker (\ref{4-5}), (\ref{5-3}) in a dual hopping
environment (\ref{4-4a}).  This exact result allows for the exact
calculation of the density in specific environments, but can also be
used for extremely accurate numerical calculations of the density for
any fixed inhomogeneous hopping environment or for disordered systems
with subdiffusive logarithmic behaviour. The approach has a
straightforward extension to the calculation of correlation
functions. 

(iv) For disordered environments which are on average translationally
and reflection invariant and which lead to an asymptotic factorization
of the two-particle conditional probabilities,
the density  at time $t$ with a random initial state is equal to the
return probability (\ref{12}) of a single particle. An important open problem 
is the derivation of conditions on disorder distributions under which 
factorization holds. For such distributions 
(\ref{12}) is an exact asymptotic result, and, as we would like to point
out, consistent with the Smoluchovsky approach if the return
probability is proportional to $1/\sqrt{D_{eff}t}$. This may be a hint
under which circumstances the Smoluchovsky treatment is adequate for
the calculation of the density.  Since at late times particles are
separated (on average) by a distance $1/\rho \to \infty$, one would
not expect a finite reaction rate or short-range interactions between
particles to change this asymptotic behaviour. It is interesting to
note that (\ref{12}) is also consistent with the renormalization group
treatment of DLPA with disorder \cite{Card97}. 
Therefore we are confident that (\ref{12}) holds for a broad class
of disorder distributions.

\section*{Acknowledgments}

G.M.S. would like to thank the Department of Physics, University of Oxford,
for kind hospitality and for providing a
stimulating environment. In particular, we thank M.J.E. Richardson
and J.L. Cardy for communicating preliminary results obtained
from the RG treatment of DLPA.

\newpage

\begin{center}
{\bf Figure Captions}
\end{center}

\noindent FIG.\ 1. Factorization of one particle
condition-probabilities after the disorder average. $R(t)$ vs. $t$ is
shown in a double logarithmic plot. The line has the slope $-0.49$ and
is a fit of the first 7 points which are produced by exact numerical
calculations of the master equation where the jump-probabilities $b_k$
were taken randomly from a uniform distribution between $0.05 < b_k
\leq 1/2$.  

\vskip 0.5cm

\noindent FIG.\ 2. Factorization of one particle
condition-probabilities after the disorder average. $R(t)$
vs. $log(t)$. The points are produced as in Fig 1, but the
jump-probabilities $b_k$ were taken randomly from a uniform
distribution between $0 < b_k \leq 1/2$. 
\vskip 0.5cm


\begin{references}
\bibitem{oddsec}
The sector with an
odd number of particles can be transformed in a similar way using
${\cal D}_- $. Some care needs to be taken in the treatment
of the boundary term.
\end{references}
\end{document}